IT UNIVERSITY OF COPENHAGEN

# Evaluating GNNs for Success Prediction in Artist Collaboration Networks

IT University of Copenhagen
Denmark

**Wiktor Dowgiałło**

wiktor.pedrycz@gmail.com

Supervisor: Michele Coscia

# Abstract

As the music industry becomes an increasingly collaborative effort, understanding the underlying structures of the artist network has become a focal point in cultural data analytics. This study expands on the previous analyses of the Italian and Danish networks by introducing a novel dataset of the Polish music scene. By utilizing methodologies used in the prior studies, this work enables a direct comparison between three distinct European music landscapes and allows to merged the created networks into one. Furthermore, this research introduces a framework to test the efficacy of Graph Neural Networks (GNNs) for artist popularity predictions based on the metadata and the position in the network. The statistical analysis revealed that the Polish and tri-national network exhibit similar properties and clustering behaviours, consistent with prior models. An evaluation of the predictive architectures reveals that while GNN models achieve a comparable F1-macro scores to the Multilayer Perceptron (MLP) in specific cases however the MLP remains a superior model regarding the success metric. The results suggest that internal node features – such as genre and label affiliation might carry more predictive capabilities than the topology of the network. The higher performance of the GNN models in the tri-national network might also suggests that the relational features become more informative when the network spans multiple linguistic and geographic boundaries, with the GNNs potentially capturing complex 'bridge' structures between the merged networks.

# 1.Introduction

While musical collaboration was once dictated by proximity and connections, rapid technological advancements fundamentally restructured the creative landscape. As the frequency of collaborations in music increased, so did the complexity of the relationships between artists. With network analysis being the most prominent method of studying cultural data analytics, the use of collaboration links has become a standard approach to model the music industry [1-5]. Other network-based studies in the music industry focus on diverse areas: prioritizing music production as an import-export network between countries [6], researching how consumption patterns define genres [7], and exploring how music listeners affect each other [8–10].

New studies [11-12] have created Italian and Danish networks making use of a range of innovative distance measures [13-16] to map out the musical landscape of those countries. The papers focus on:

1. Discovery of eras in music and genres defining them.

2. The concentration of genres and the spreading of genre trends along the network.
3. Analysis of similarities and differences between the two created networks and the comparison of network statistics.

Some other papers focus on the analysis of different music industries in countries such as those in United States, Ireland, France, Canada, Germany, and Spain [17-22]. None of these papers cover the chosen area of interest making the Italian and Danish networks [11-12] unique in their data model and application of tools.

Building upon the foundations of the Italian [11] and Danish [12] networks, this paper seeks to expand on the previous analysis by developing a novel Polish artist network. By replicating the methodologies used in the previous studies, this research allows for integration of all three datasets into a singular, unified network. The merged network is compared throughout the study with the Polish network to identify the variations between localized artist clusters and the macro-level structures of the European music landscape.

Furthermore, this study evaluates the efficacy of Graph Neural Networks (GNNs) for predictive modelling within musical artist ecosystems utilizing the acquired data. While prior studies confirm the high performance of Graph Neural Networks (GNNs) in artist networks for tasks such as determining similarity [22-24], music classification [25] & predicting collaboration links [26-27], their application in forecasting artist popularity remains significantly understudied. Whereas existing studies tend to focus on song-level success [28] or artist success predictions using classical machine learning methods [29], this paper investigates the extent to which artist's network position can predict their popularity based on the unique features obtained in this study.

The analysis in this paper utilizes both the newly constructed Polish network and the merged tri-national dataset to examine the following:

1. Statistical, Structural and Temporal Dynamics: A comparative analysis of the Polish and merged network. This allows for a direct comparison with previously established studies [11-12]
2. Predictive modelling via GNNs: An evaluation of the viability of Graph Neural Network architecture for the forecasting of artists success metric and benchmarking it against classical deep learning methods.

With the previous works focusing on two countries, the research community could benefit greatly from a more generalized comparison between three countries from different European language groups. By selecting representative countries from the Romance, Germanic, and Slavic linguistic families, this study can more effectively explore the cultural and structural nuances that influence collaborative patterns across Europe.

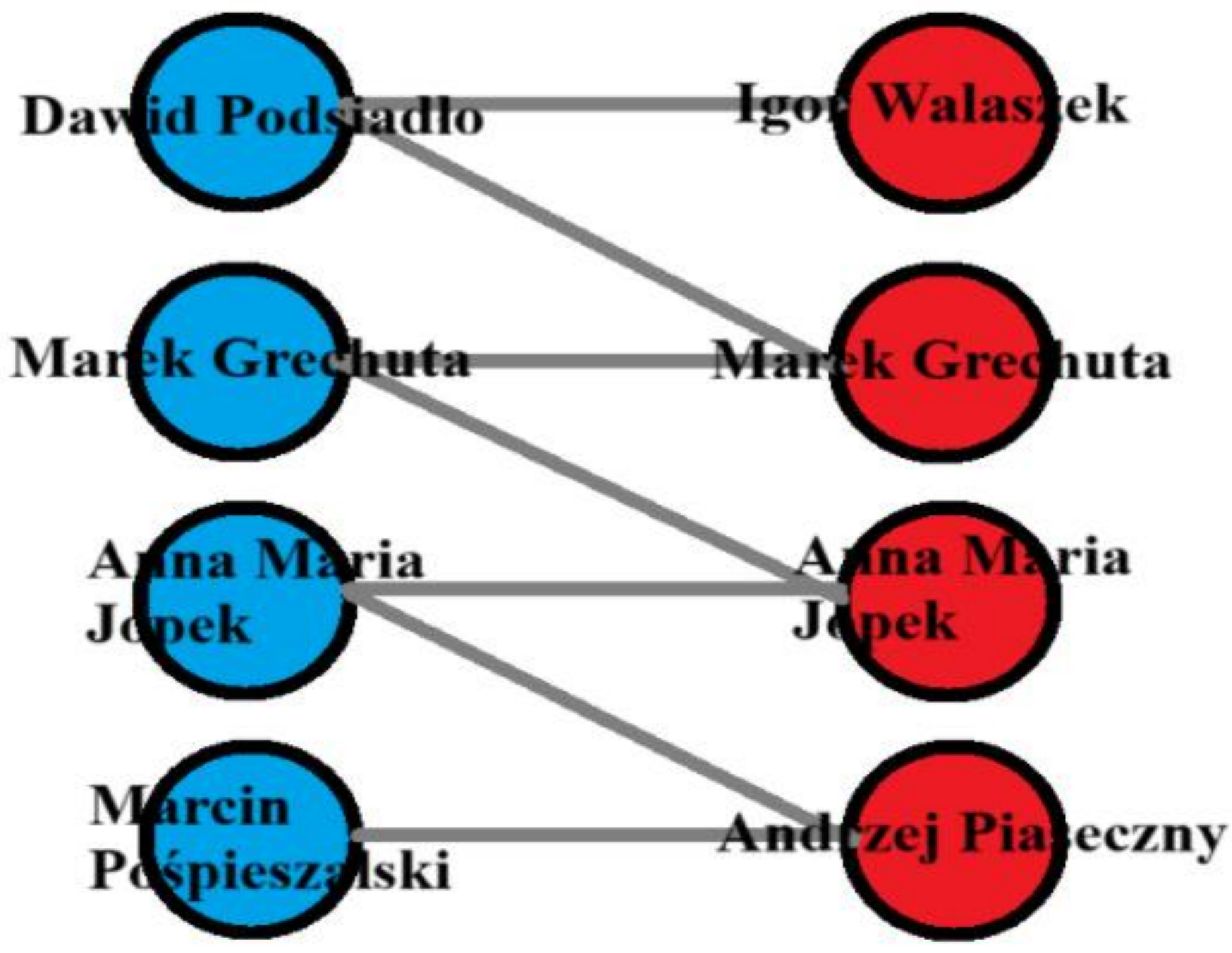

**Figure 1.** An example of the bipartite network. The nodes on the left are artists and the nodes on the right are bands.

The primarily used dataset in this analysis, containing the newly developed Polish network can be found here: [https://doi.org/10.5281/zenodo.16870759](https://doi.org/10.5281/zenodo.16870759).

# 2. Data & Methodologies

## 2.1 Data Collection

The Polish artist and band network was entirely constructed using ‘Discogs’[1], utilizing metadata collected between February and October 2025. To ensure the network is representative of the Polish music scene, artists were filtered by Poland as the country and sorted by collection frequency. The total amount of bands collected was 658, so that the proportion of bands per number of releases displayed on the website is equivalent to the proportion of the Italian and Danish networks. To guarantee the collected data reflects the variety of years, the dataset contains all artists performing from 1900-1950. The remaining years were sampled proportionally based on the release volume of the 30 most prevalent music styles in Poland on the platform. From this stratified sample, the top chosen artists in each genre were selected for inclusion. During this process, entries incorrectly attributed to Poland by the platform were manually removed.

[1] https://www.discogs.com/

| Band name | Genre | | Year | | Label ID | |
|---|---|---|---|---|---|---|
| | Abstract | Ambient | 2010 | 2011 | 140264 | 13929 |
| Noon | 6 | 5 | 0 | 1 | 0 | 0 |
| Jacek Sienkiewicz | 7 | 4 | 0 | 0 | 3 | 0 |
| Kury | 1 | 0 | 0 | 0 | 0 | 0 |
| Job Karma | 0 | 19 | 1 | 0 | 0 | 2 |

**Table 1**: An example of two categories from each attribute type for 4 chosen bands. The numbers are not normalized.

For each unique release, the performer data was extracted via a semi-automatic pipeline. This involved a parsing script which detected the performer information and retained only the individuals with core roles (e.g. musicians, performers) based on the release information. The script saved the filtered artists into a file of Tab-Separated Values (TSV) format. The resulting dataset consisted of four fields: Artist, Band, Year and Role. To ensure the uniqueness of each row, only one chosen album from a band per year has been parsed.

## 2.2 Data Model

The network models the relationship among Polish artists and bands. The data model, cleaning and pre-processing closely replicate those used in the previous works of the Italian and Danish networks [11-12] and aim to create a comparable structure.

The data model used is a temporal bipartite network with structure G = (V1, V2, E). The nodes in the first class V1 are artists, real physical people. The nodes in the second class are bands, identified by their name. An edge connects an artist to a band in a specific year. Solo artists are also considered as bands, and they are logically different from the artists with the same name. For instance, there is an artist Marek Grechuta and a band Marek Grechuta. The artist Marek Grechuta is the physical singer while the band is the collective of artists who released music under the name Marek Grechuta over the years. The artist performs for the band of the same name, but they are not the same entity. Figure 1 show an example of the structure of the bipartite network.

Each band node in the dataset also holds multiple attributes classes: genre, year and label. Genres and labels are collected from the Discogs XML dump of releases and labels whereas the year is taken from the previously processed TSV file. The attributes are stored in a frequency matrix format, where each number represents the total amount of releases associated with a given genre, year and label category. The year matrix is binary due to the previously described method of parsing one album per year from each band. The structure is presented in Table 1.

The labels are processed further to address the fragmentation of record labels. In the Discogs dataset, a single label entity is represented by many sub-labels or parent labels which artificially increase the number of unique labels. To resolve this issue, a label disambiguation is performed. The sub-labels and parent labels are mapped to a single canonical label ID. Where a clear hierarchy exists, the root label is selected. In other cases, the canonical label is determined by the volume of production.

## 2.3 Data Processing

As mentioned in the previous steps, the created network consists of only the Polish bands. A band is considered Polish if it has released a significant proportion of their discography in Poland. This criterion ensures that the network stays predominantly Polish while accounting for overlap with other countries.

Following closely the established methodologies of the Italian and Danish networks [11-12], an entity resolution process is implemented to identify and merge duplicate artist profiles. For this task, the Ratcliff-Obershelp [30] string distance is computed for all pairs of artists. When the similarity metric indicates a high probability of a match, entries were verified and standardized to a single name upon manual confirmation. This process helps improve the network integrity and ensures collaboration links are attributed to the correct individuals.

Finally, a network projection is performed to transform the bipartite artist-band network into a unipartite band-band network. A noise-corrected backboning algorithm [31] is applied to only keep the statistically significant band-band connections. This step helps mitigate the impact of unresolved pseudonyms or homonyms that may have bypassed the initial cleaning stage. That is because of mistakes caused by incorrect names that change the edge weight. After manual thresholding only significant weights remain.

In order to determine the edge weights in the projection, a temporal overlap approach is utilized as described the formula:

$$w_{b_1,b_2} = \sum_{a \in b_1 \cap b_2} \min\left(w_{a,b_1}, w_{a,b_2}\right)$$

For each artist that bands b1 and b2 share, a minimum number of years during which a played for either b1 or b2. This ensures that edges between bands are stronger for artists who contributed a significant amount to both bands history. For example, the entity Maryla Rodowicz and Smolik have a score of 23. This means that from all artists that collaborated with both of the bands, the lowest amount of releases that each artist contributed to any of the two bands is taken and summed, equalling 23.

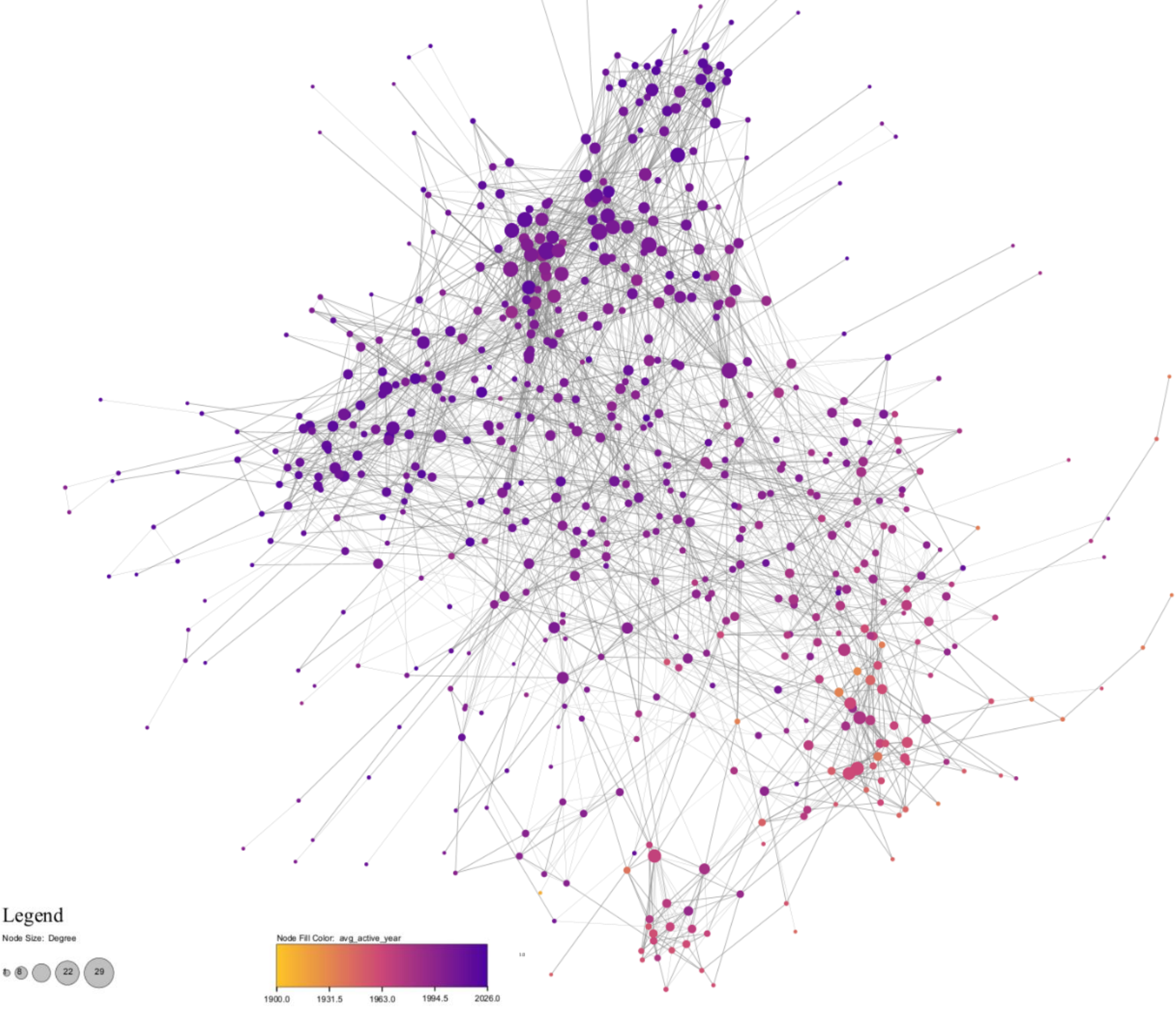

**Figure 2**: The Polish network visualized using the force directed layout. The degree is indicated by the size of the node and the average year of activity is indicated by the colour

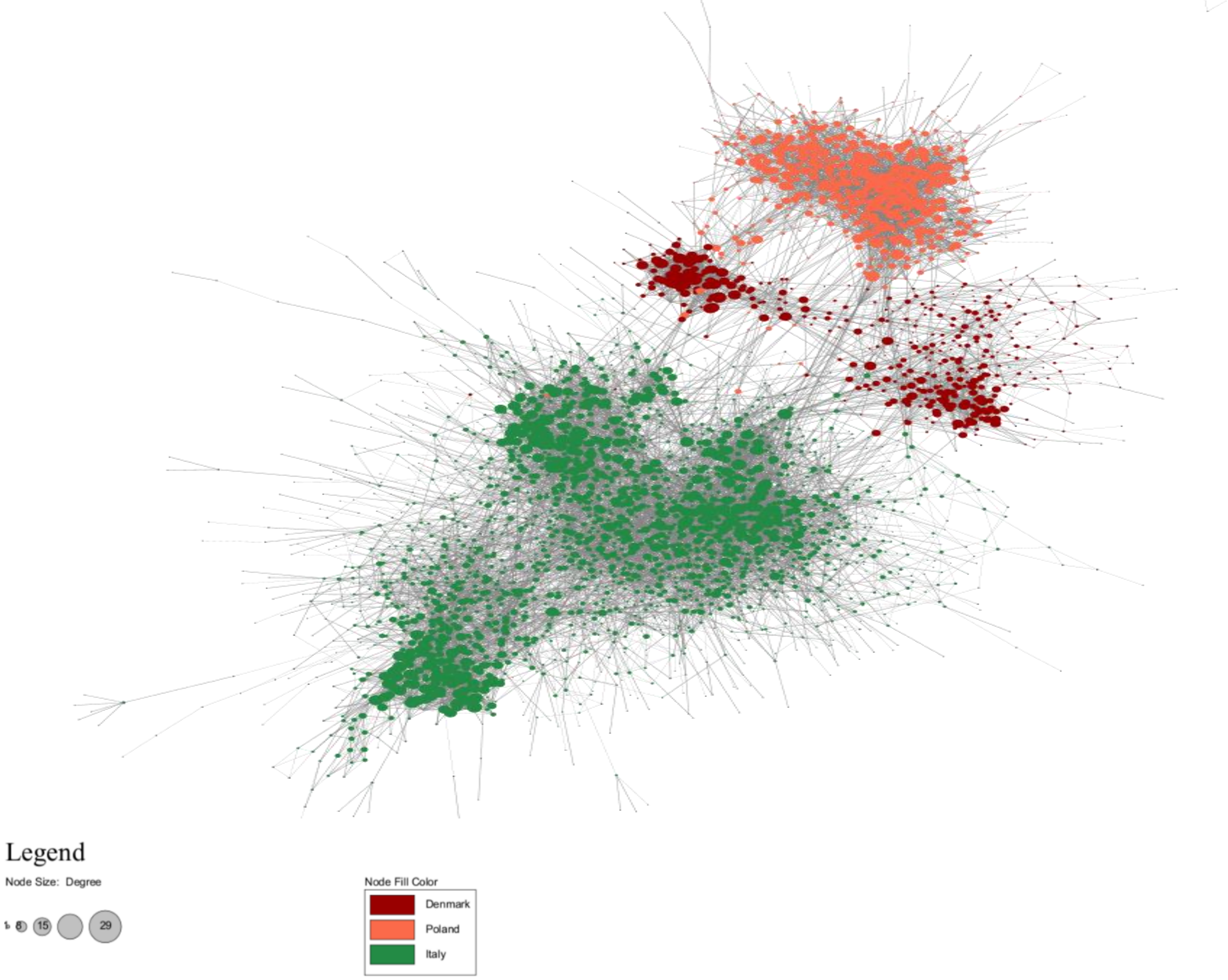

**Figure 3**: The merged network visualized using the force directed layout. The degree is indicated by the size of the node and the country of origin is indicated by the colour

# 3. Network Analysis

## 3.1 Network Visualizations

To gain a deeper understanding of the newly formed networks, the networks are visualized. Visualizing the networks allows for a qualitative assessment of the topological structures and a simple identification of the core and periphery of the network.

In the visualization of the Polish network (Figure 2), the nodes are scaled according to their degree centrality, where the size corresponds to the amount of connections a band has. The temporal evolution of the network is denoted by the colour, with the gradient representing the average year of activity of a band. This encoding method displays the development of the network over time while highlighting key hubs.

For the merged tri-national network (Figure 3), the node size remains consistent with degree centrality. However, the colour gradient is switched to categorical scheme, representing the country of origin. This method enables a directed comparison of the three networks, illustrating clearly the density of each cluster, and the intersections between them.

**Poland**

| Variable | Value |
|---|---|
| # Bands | 658 |
| # Artists | 12,512 |
| # Edges | 26,723 |
| # Years | 99 |
| # Band AVG Year | 5.3 |

**Merged**

| Variable | Value |
|---|---|
| # Bands | 3,519 |
| # Artists | 49,978 |
| # Edges | 154,241 |
| # Years | 123 |
| # Band AVG Year | 6.7 |

**Table 2**: Summary statistics for the bipartite networks of Poland and the tri-national.

**Poland**

| Variable | Artist | Band |
|---|---|---|
| # Nodes | 12,506 | 641 |
| # Edges | 27,312 | 2,868 |
| Avg Deg | 4.4 | 8.9 |
| Density | 0.0003 | 0.0140 |
| Clustering | 0.5216 | 0.2770 |
| Modularity | 0.9838 | 0.5992 |

**Merged**

| Variable | Artist | Band |
|---|---|---|
| # Nodes | 50,016 | 3,502 |
| # Edges | 194,270 | 11,351 |
| Avg Deg | 7.8 | 6.5 |
| Density | 0.0002 | 0.0019 |
| Clustering | 0.5491 | 0.3585 |
| Modularity | 0.9084 | 0.8508 |

**Table 3**: Summary statistics for the projected networks of Poland and the tri-national. The node count differs from Table 2 due to backboning and isolated nodes.

## 3.2 Network Statistics

To enable a formal quantitate comparison of the Polish and the tri-national merged network to the previously established networks [11-12], the basic statistics for the networks are computed. The statistics for the bipartite networks are presented in Table 2 and the projected networks in Table 3.

The main difference between the networks is the size, with the tri-national network containing more than 5 times the amount of Polish bands, the largest contributor being Italy (2447 nodes) followed by Poland (641 nodes) and Denmark (415 nodes). Due to the increase in number of nodes in the merged network, the density is lower than in the Polish network. This is understandable for growing networks considering that when a pool of artists expands, a chance of collaboration between two bands decreases.

The average degree for tri-national network bands is lower than for the Polish bands however it is higher for the artists. This pattern follows from the previous comparison of the Italian and Danish networks [12] suggesting that it might be the massively represented Italian music scene with high individual artist mobility causing the shift. The Figures 4 and 5 show the degree distributions for the bipartite networks and all the unipartite projections – by artist and band. The general dynamics of the networks align, slightly differing in scale.

The temporal activities of the networks do not differ significantly either as it can be seen in Figure 6. The size of the network being again the only noticeable contrast.

Statistics not fully dependant on the size of the network such as the clustering coefficient and modularity show consistency in the structure regardless of the scale. Specifically, the high clustering coefficient for artists across both networks indicates the tendency of musicians to organize into tight-knit cliques. The metric is however lower for the band projected networks suggesting that bands tend to form more distinct and specialized entities. While high modularity for the Polish network imply a scene characterized by sub-communities, the disparity of the metric between the band projection of both networks highlights a shift from the local based fragmentation to geographic isolation. This confirms that the geographic proximity remains the primary driver of the architecture as it can be seen in the tri-national network (Figure 3).

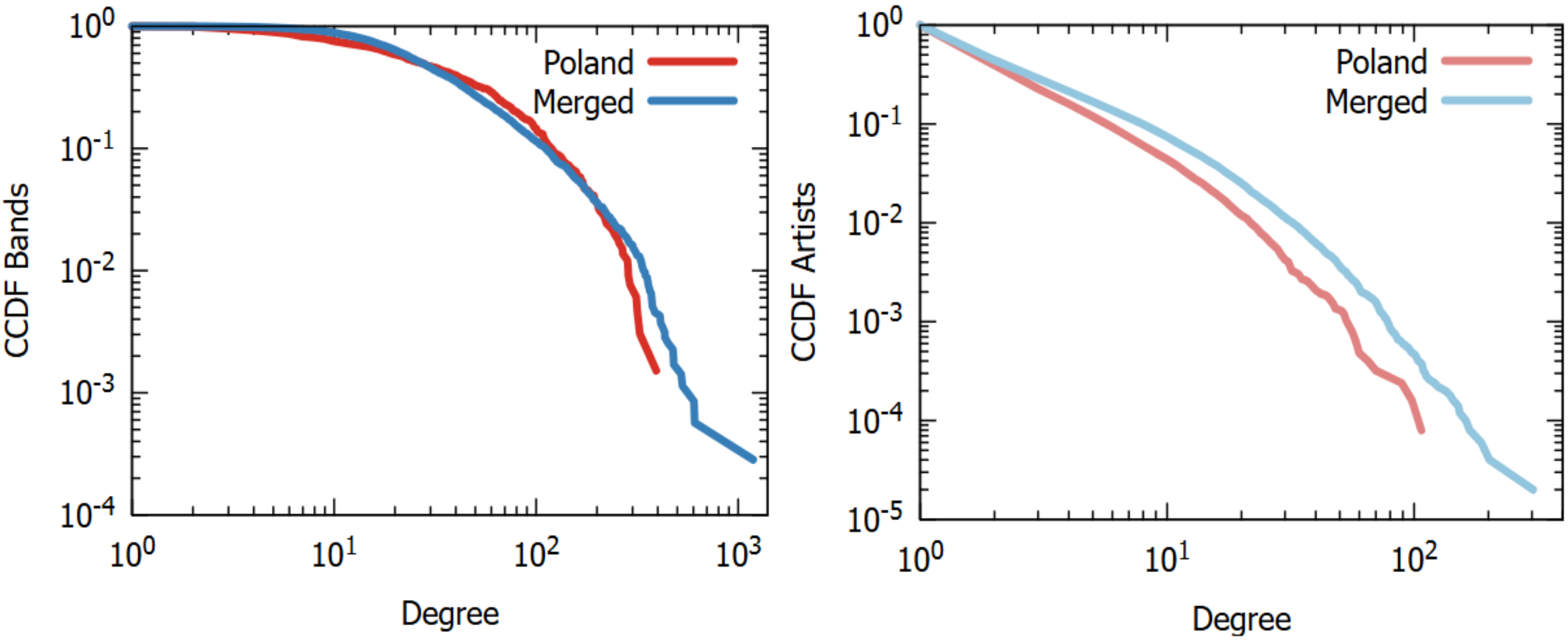

**Figure 4**: Complement of the cumulative degree distribution in the bipartite network for (right) bands and (right) artists. Tri-national in blue and Polish in red.

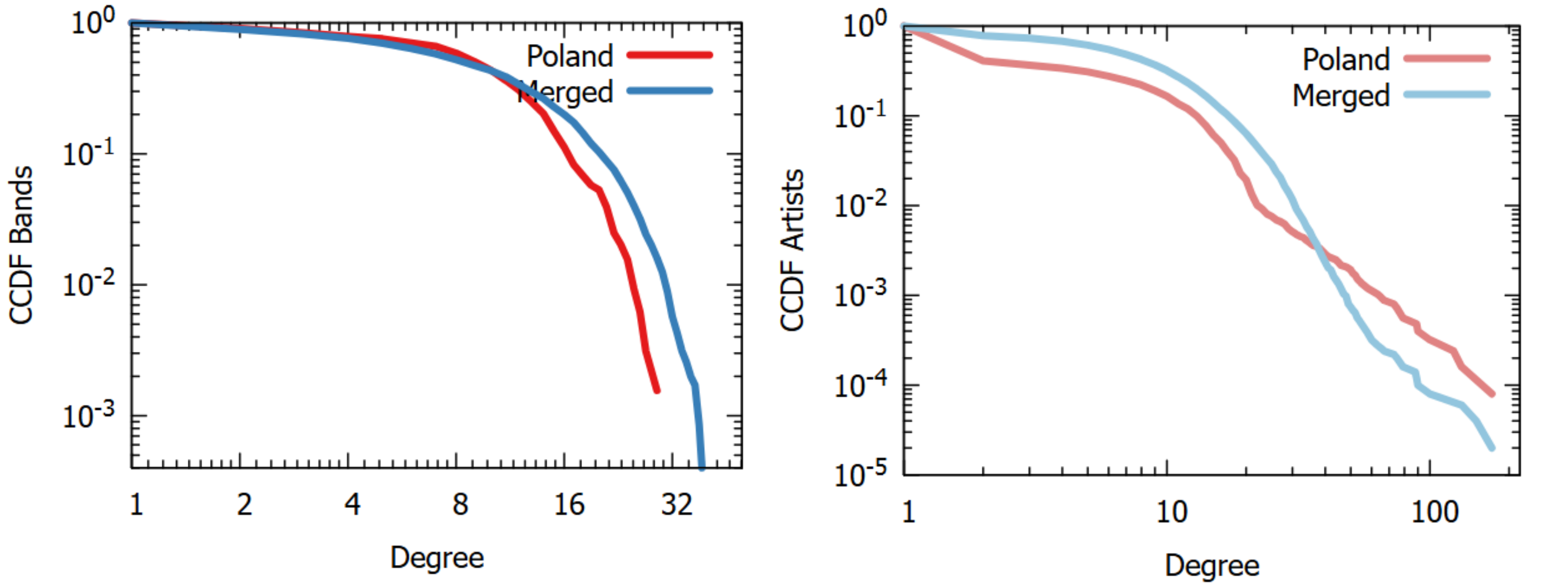

**Figure 5**: Complement of the cumulative degree distribution in the unipartite networks for (left) bands and (right) artists. Tri-national in blue and Polish in red.

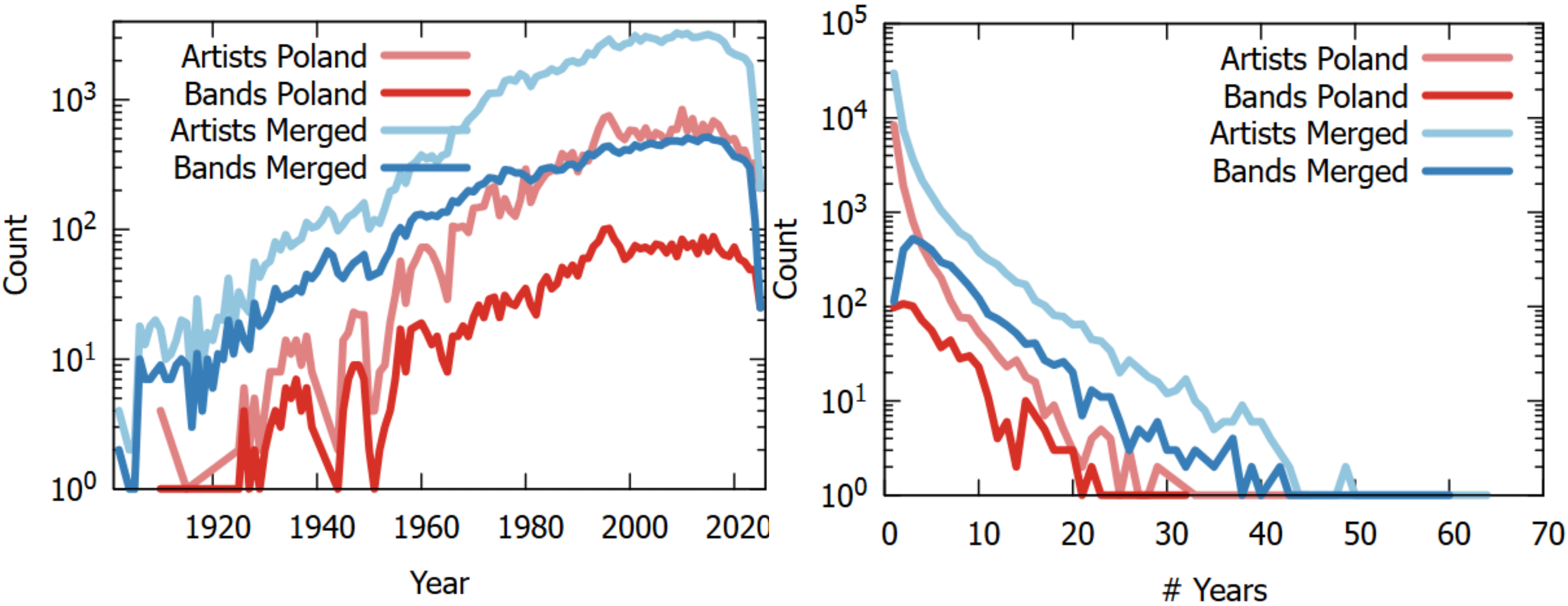

**Figure 6**: (Left) Number of distinct bands and artists (y axis) per year (x axis). (Right) Number of bands and artists (y axis) with a given number of activity (x axis). In both subfigures, tri-national is in blue and Poland is in red.

## 3.3 Band connections

In order to get a better understanding of the underlying reason why bands connect each other in the projection, a simple model consistent with previous works [11-12] is utilized. A linear regression is implemented to test if an edge is significant after backboning. The binary variable Y is introduced to indicate if b1 and b2 have a statistically significant connections. Three metrics are used to measure the band-band similarity:

- G - Genre similarity, calculated by applying logarithm to the weighted overlap coefficient, same as the one used for weights in the band projection. The number of records a band published in a given genre is normalized so that the sum across all genres adds up to one. This helps prevents generalized bands having an inflated similarity.
- T – Temporal similarity, taken by taking the logarithm of the years in which the bands released a record.
- L – Label similarity, computed using the same logic as the genre similarity.

The results of the regression can be seen in Figure 7, which shows the amount of variance we can explain in the band-band network edge existence probability. The results indicate that the networks are similar, with the majority of variance remaining unexplained by the three variables. The label contributes the most to the explanation and the year the least. This is different than the finding in the previous comparison [12] in which the label contributed the least and the genre was the most significant. This discrepancy suggests that the labels in the Polish and tri-national network serve as the primary catalyst for collaboration, hinting at a higher organizational influence in the computed networks.

## 3.3 Key nodes

In the context of high modularity, identifying key nodes becomes a crucial part of getting a better understanding of the network topology. Determining which nodes serve as bridges (betweenness centrality), anchors (degree centrality) and influencers (PageRank) provides valuable information about the network dynamics.

The Table 4 shows the top 10 nodes ranked by each method for the Polish and tri-national merged network. Italian nodes emerges as the networks core, dominating both degree centrality and PageRank which can be again attributed to its volume – the absolute highest degree centrality belongs to a Polish node.

The betweenness centrality gives more insight into the key nodes connecting the three networks together. The metric was approximated using 500 samples of source nodes for faster computation. Half of all entries are Italian and half are Polish. Most of the top entries are Polish which suggest that the Polish network has a key role as the connective tissue of the tri-national musical landscape. No Danish entries were in the top 10 of any metric.

# 4. Graph Neural Network

Traditional machine learning models, such as the Multilayer Perceptron (MLP) process the data points as isolated entities, making generalized predictions that often fail to capture the nuances of complex dependencies within a dataset. In the music industry, this approach could fail to notice that a band's success is often intertwined with their collaborators. The structural properties of the bipartite network and the created projections suggest that a Graph Neural Network (GNN) is the more appropriate architecture for this analysis, shifting the focus to relational context for band's success forecasting.

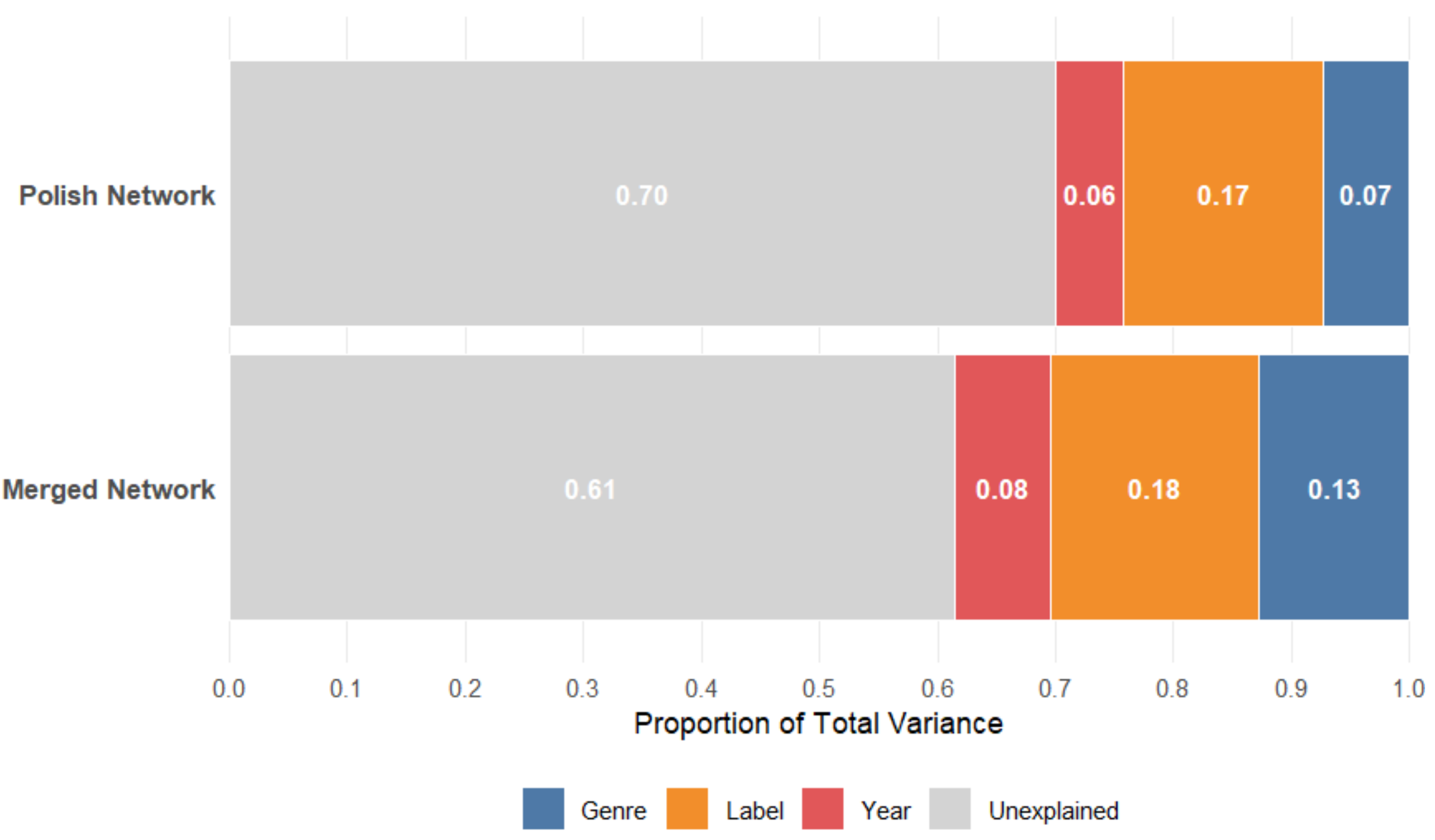

**Figure 7**: The amount of variance we can explain in the band-band edge existence probability for Polish and tri-national network. Colours show the variance predicted by different similarities: genre (blue), label (orange), and temporal (red). Gray shows the residual, the amount of variance unexplained by the three variables.

| Rank | Degree | | Betweenness | | PageRank | |
|---|---|---|---|---|---|---|
| | Poland | Merged | Poland | Merged | Poland | Merged |
| 1 | Eye For An Eye (29) | Vader (39, POL) | Quo Vadis | Blindead (POL) | Eye For An Eye | Death SS (ITA) |
| 2 | Trauma (27) | Genius (38, ITA) | Witold Rowicki | Temperance (ITA) | Witold Rowicki | Temperance (ITA) |
| 3 | Vader (26) | Temperance (38, ITA) | Żywiołak | P. K. Drazek (POL) | Quo Vadis | Odyssea (ITA) |
| 4 | Quo Vadis (26) | Beniamino Gigli (37, ITA) | Kury | Monopol (POL) | Trauma | Genius (ITA) |
| 5 | Guernica Y Luno (25) | Odyssea (37, ITA) | Azarath | Elvenking (ITA) | Azarath | Francesco Paladino (ITA) |
| 6 | Lipek (25) | Tino Vailati (37, ITA) | Magdalena Woźniak | Hate (POL) | Maria Koterbska | Beniamino Gigli (ITA) |
| 7 | Azarath (24) | Oscar Carboni (36, ITA) | Anna Dabrowska | Stormo (ITA) | Lipek | DJ Dado (ITA) |
| 8 | Deus Mortem (24) | Marnero (35, ITA) | Blindead | Robertino (ITA) | Vader | Doctor's Cat (ITA) |
| 9 | Riverside (24) | Death SS (35, ITA) | Boys | Genius (ITA) | Guernica Y Luno | Marnero (ITA) |
| 10 | Hate (24) | Stormo (34, ITA) | Kalwi & Remi | Marek Biliński (POL) | Riverside | Labyrinth (ITA) |

**Table 4**: Top 10 bands ranked by degree centrality, betweenness centrality and PageRank for the Polish and tri-national merged network. The degree section displays the degrees next to the entry. The merged network also displays the 3 letter abbreviation of the bands country of origin.

## 4.1 GNN vs. MLP architecture

The main distinction between traditional neural networks and GNNs is the treatment of data dependency. Classic architectures, such as MLP operate on the assumption of independent and identical data (i.i.d.) distribution, treating each band as isolated vectors. While effective for processing fixed features, it fails to capture the topological structure of datasets, such as the band collaboration network.

The GNNs on the other hand, does not operate on the i.i.d. assumption and leverages the structure of the network instead. It utilizes the pairwise message passing framework to exchange information with nodes' neighbours. Rather than relying solely on the nodes own features, GNN incorporates the features of the adjacent nodes, modelling how attributes propagate through a network. This transition allows the model to capture structural aspects that traditional models ignore.

## 4.2 Preprocessing

In order to prepare the data for machine learning methods, the data noise in the music industry needs to be cleaned and the data needs to be processed.

To acquire the success feature needed for the predictions, the releases from Discogs are processed using the API to collect information about the 'wants' and 'haves' of each release. This metrics represent the number of users who added a given release to their 'Wantlist' and their physical collection on Discogs respectively.

Then the files with disambiguated labels and success metrics are run through an algorithm together with the Discogs releases and artists files to create a file with information about each release (release id, canonical band id, labels, genres, styles, year, origin, have and want). The process also ensures that the attributes are cleaned and an artist who released an album under pseudonym or alias, gets the credit to their canonical id.

## 4.3 Tensor creation

Once the file has been created, the data is transformed into a format compatible with deep learning frameworks. The process constructs a tensor object including the features, target labels & the network topology.

To ensure the flow of information, a bidirectional adjacency tensor is constructed. All edges in the band-band backbone are mirrored and the backbone weights are preserved to weight the influence of collaborators. Due to the high dimensionality of the categorical data with thousands of label and genre tags, the node feature matrix is constructed as a compressed sparse row tensor.

The target labels are prepared for the specific prediction tasks. For the success metric, a logarithm transformation of the 'have' and 'want' aggregate counts is computed. This transformation was introduced with the original goal of performing regression on the data however it will not have effect on the later computed binarized success metric. Other target labels are created for country (only for the tri-national network), genre, year and label, however for the purpose of this research, only the country and success prediction task will be considered. The features used for the success metric and country predictions are genres, years and labels.

In order to make the model generalizable, the dataset is partitioned into training (70%), validation (15%) and testing (15%) sets. The sets are split utilizing stratified sampling based on the target distribution (or bins for success). This ensures that each subset is representative of the general dataset, preventing the model from becoming biased towards one label.

## 4.4 Setup

To evaluate the predictive power of the Graph Neural Networks, they are compared to the traditional architecture, to establish a benchmark. The architectures considered for this research are Multilayer Perceptron (MLP), Graph Convolutional Network (GCN) [33] and Graph Attention Network (GAT) [34]. The setup evaluates two primary tasks – country classification and success binary classification, the transformed popularity metric using the median threshold. The first task was introduced with the assumption that GNNs should perform better on a task such as country classification due to the high modularity of the band projection in the tri-national network.

The architectures of the models were chosen manually as follows:

- Multilayer Perceptron (MLP) - Feed-forward architecture consisting of three fully connected layers with batch normalization and ReLU activation functions. Uses kaiming normal initialization for ReLU to ensure that the signal does not disappear, leading to dead neurons. A high dropout rate of 0.5 is applied between layers to prevent overfitting.
- Graph Convolution Network (GCN) – Graph Neural Network architecture with three convolution layers aggregating the neighbouring node's features and normalizing the average, before combining them with its own features. Similar to MLP, uses ReLU activation function with batch normalization and the same dropout rate.
- Graph Attention Network (GAT) – Also a GNNs architecture, with attention layers learning the attention coefficients for every edge

instead of averaging the like in the GCN. In addition to the standard dropout with the same value it utilizes a 0.2 dropout on the attention weights to prevent overreliance on one neighbouring node. The first layer uses 4 heads which are concatenated in the first two layers and averaged in the final one. Same as the previously listed architectures, it also uses ReLU and batch normalization.

The loss function used is the cross entropy loss which measures the performance of classification models outputting values between 0 and 1, maximizing the probability of the correct category by penalizing the confidence of mistakes. The training process includes a patience based early stopping mechanisms utilizing the validation set. The model monitors the validation loss and restores the best weights found during training. Every model experiment is run across 5 different seeds. This ensures that the results are more a more accurate representation of the realty and also allows to report on the standard deviation, providing insight into the model stability.

# 5. Results & Discussion

The results across the tri-national merged network and the Polish network are compared on two main metrics - accuracy and the F1-macro. Both networks perform the success classification task, with the merged network additionally performing the country classification. The results are presented in Table 5.

## 5.1 Country Classification

In the country of origin predictive task, the GCN and MLP performed nearly identical, with MLP achieving a slightly higher accuracy and GCN achieving a higher F1-macro. The high performance of the MLP suggests that a band's country is mostly encoded within its local features such as labels and genre. Since labels tend to operate within national borders and genres can be specific to regions or countries, the attributes can act as strong independent predictors.

The GCN significantly outperformed the GAT in both accuracy and F1-macro. This indicates that country affiliation is a property which benefits more from information from all neighbours rather than selective attention. The GAT's lower performance could be attributed to its attention mechanism scrambling a clearly defined signal.

## 5.2 Success Prediction

The prediction of success binarized at median, the main focus of the study, provides the largest contrast between the tri-national and regional datasets.

| Dataset / Task | Model | Accuracy | | F1-Macro | |
|---|---|---|---|---|---|
| | | Mean | Std | Mean | Std |
| **Merged Dataset** | | | | | |
| Country (Class.) | GAT | 0.8583 | 0.0055 | 0.7017 | 0.0100 |
| | GCN | 0.9129 | 0.0143 | **0.8514** | 0.0285 |
| | MLP | **0.9132** | 0.0053 | 0.8463 | 0.0080 |
| Success (Class.) | GAT | 0.8418 | 0.0086 | 0.8412 | 0.0090 |
| | GCN | 0.8430 | 0.0083 | 0.8428 | 0.0086 |
| | MLP | **0.8603** | 0.0067 | **0.8598** | 0.0069 |
| **Polish Dataset** | | | | | |
| Success (Class.) | GAT | 0.6761 | 0.0209 | 0.6753 | 0.0211 |
| | GCN | 0.6326 | 0.0049 | 0.6213 | 0.0061 |
| | MLP | **0.8717** | 0.0161 | **0.8706** | 0.0165 |

**Table 5**: Results of the test set of two different tasks across three models and two datasets. Aggregated result of 5 runs on different seeds with mean and standard deviation being reported.

On the merged dataset, the MLP once again achieved the highest result, slightly outperforming the GCN and GAT. This suggests that the used predictive features (label, genre and year) are dominant, making additional information about the structure of the network be considered noise, degrading the GNNs' performance.

In the Polish network, the MLP maintains its high accuracy and F1-macro, however the results are vastly different from the tri-national dataset. The difference between the traditional machine learning method and the network based is significantly higher, indicating that the previously noticed nuance is even more prevalent in the Polish network. With the GAT outperforming the GCN, it becomes clearer that in smaller and more specific musical landscapes, the GAT's ability to selectively weight the importance of neighbours is more effective than GCN's simple averaging.

## 5.3 Discussion of Findings

With MLP's consistently high performance, it becomes clear that in this dataset, the structure of the networks does not provide higher quality information about the data. This indicates that the preprocessing phases were able to create a strong signal with the capability to capture nuances usually found in network structures. It also shows that features such as labels and genres are far more specific to countries and regions.

The disparity between GCN's and GAT's performance in the country classification task and the Polish success prediction task shows that the geographic origin is a collective signal whereas the Polish musical scene is more driven by quality of connections. This is understandable considering that most artists tend to collaborate with their own countrymen and popularity is far more influenced by connections to important bands.

# 6. Limitations

Several limitations regarding the data structure, data collection and methodologies must be acknowledged.

Due to the constraints of automatic collection such as lack of verification or formatting issues, the data collections has been performed semi-automatically. This limited the amount of possible bands to encode and resulted in a lower but more truthful coverage of the Polish musical scene. While the data was cleaned thoroughly, it may still contain human-generated inaccuracies and mistakes.

Another limitation is the reliance on Discogs as the primary data source for the creation of the success metric. Since the website is primarily used by physical media enthusiasts, the 'have' and 'want' metric are biased towards the preference of those users. Genres with a strong CD or vinyl cultures like Heavy Metal or Electronic music tend to be over-represented in the dataset in regards to the popularity.

# 7. Conclusion

This study contributes to the field of cultural data analytics, particularly focusing on dynamics and structure of musical artist collaborations. It expands on the previously created networks of Italy and Denmark [11-12] by introducing a novel Polish dataset. The created datasets are combined into a tri-national network which is then compared with the Polish network. They are compared using metrics and visualizations previously used on Danish-Italian network comparison [12], applying same methodologies. Furthermore, the datasets are used to test the efficacy of Graph Neural Network (GNN) predictions for country of origin predictions and the created success metric.

The comparison of the networks shows that the networks share similar structural properties, with the largest difference being the size. Despite the significant variance in node count, the degree distributions and temporal activities align very closely across the networks. The band collaboration links of the tri-national and Polish networks are explained by similar proportions, however differing slightly from the previous study [12].

The implementation of GNNs reveals that while graph based machine learning methods can successfully utilize the structure of the network for predictions, the results are very task dependent. For the merged network, the GNNs achieved close results to the Multilayer Perceptron in both country classification and success prediction tasks. In contrast, the success predictions for the Polish dataset show vastly different results in performance for all models with MLP achieving the highest accuracy and F1-Macro, followed by GAT and GCN. The disparities between the datasets suggest that the relational features are more informative when the network spans multiple linguistic and geographic boundaries.

This research highlights the need for future studies in this area. While the addition of the Polish networks provides a crucial addition to the Italian and Danish datasets, it underlines the need for a further expansion to fully map the European musical landscape. Integrating more countries would help determine if the structural properties hold across different cultural and linguistic conditions. This would also allow for the further, more robust validation of the GNN architecture and the comparison against the MLP model to test whether the observed patterns in the tri-national and Polish network stay consistent with new additions. This field of research could also benefit more from future works focused on the in depth analysis of GNN architectures and the comparison of their performance across various metrics for the success prediction task.

# Data availability

The data is available at Zenodo at https://doi.org/ 10.5281/zenodo.16870759.

# Acknowledgements

I thank Lisandro Benetti and Daniel Fejerskov-Quist for their help with the collection of the Danish network and Michele Coscia for his invaluable guidance and the creation of the original Italian network.